\documentclass{elsart}
\usepackage{amssymb}
\usepackage{graphicx}
\begin{document}
\begin{frontmatter}
\title{Dihedral-angle Gaussian distribution driving protein folding\thanksref{agradecimentos}}
\author[ufpe]{P.\, H.\, Figueir\^edo\corauthref{phf}}
\ead{phf@lftc.ufpe.br}
\corauth[phf]{Corresponding author}
\author[cairu]{M.\, A.\, Moret}
\ead{moret@cairu.br}
\author[ufba]{ E.\, Nogueira Jr.}
\ead{enogue@ufba.br}
\author[ufpe]{S.\, Coutinho}
\ead{sergio@lftc.ufpe.br}
\address[ufpe]{Departamento de F\'{\i}sica,
Universidade Federal de Pernambuco,\\ CEP 50670-901, Recife,
Pernambuco, Brazil.}
\address[cairu]{Programa de Modelagem
Computacional, Funda\c{c}\~{a}o Visconde de Cairu, \\CEP 40226-900,
Salvador, Bahia, Brazil, and \\ Departamento de F\'{\i}sica,
Universidade Estadual de Feira de Santana, \\CEP 44031-460, Feira de
Santana, Bahia, Brazil.}
\address[ufba]{Instituto de F\'{\i}sica,
Universidade Federal da Bahia, \\CEP 40210-340, Salvador, Bahia,
Brazil.}
\thanks[agradecimentos]{This work received financial support from CNPq and CAPES
(Brazilian federal grant agencies), from FAPESB (Bahia state grant
agency) and from FACEPE (Pernambuco state grant agency under the
grant PRONEX EDT 0012-05.03/04).}

\begin{abstract}
The proposal of this paper is to provide  a simple angular random
walk model to build up polypeptide structures, which encompass
properties of dihedral angles of folded proteins. From this model,
structures will be built with lengths ranging from 125 up to 400
amino acids for the different fractions of secondary structure
motifs, which dihedral angles were randomly chosen according to
narrow Gaussian probability distributions. In order to measure the
fractal dimension of proteins three different cases were analyzed.
The first contained $\alpha $-helix structures only, the second
$\beta $-strands structures and the third a mix of $\alpha
$-helices and $\beta $-sheets. The behavior of proteins with
$\alpha $-helix motifs are more compacted than in other
situations. The findings herein indicate that this model describes
some structural properties of a protein and suggest that
randomness is an essential ingredient but proteins
are driven by narrow angular Gaussian probability distributions and not by
random-walk processes.
\end{abstract}

\begin{keyword}
protein structure, angular Gaussian-walk, radius of gyration, mass fractal
exponent.
\PACS{87.14.Ee, 87.15.Aa, 05.45.Df, 05.40.Jc}
\end{keyword}
\end{frontmatter}

The manner in which a protein folds from a random coil into a unique
native state in a relatively short time is one of the fundamental
puzzles of molecular biophysics. It is well accepted that a unique
native three-dimensional structure, characteristic of each protein
and determined by the sequence of its amino-acids sequence, dictates
protein functions. The folding process should involve a very complex
molecular recognition phenomenon depending on the interplay of many
relatively weak non-bonded interactions. This would leads to a huge
number of possible final conformations under conventional molecular
optimization methods based on the search for the minima of the
energy hyper-surface. This number, which should  increases with the
number of the chain's degrees of freedom, however, is severely
restricted during the real folding process, excluding relevant
portions of the energy landscapes as far as an extended or random
conformation is chosen as the initial state
\cite{dill90,wolynes95,yon97,zhou99a,zhou99b,moret98,wales99,moret01,moret02,moret05a}.
On the other hand, if the extreme limit, were considered, where a polypeptide chain departs
from its denatured state and in very relatively short period of time
finds its unique native state after searching amongst the
astronomical number of possible configurations, the simulating
process for proteins with fifty to five hundred amino acids using
approaches such as Monte Carlo and molecular dynamics, becomes
impracticable, due to the very high computation cost. Such
contradictory dynamical picture is known as \emph{Levinthal paradox}
\cite{levinthal68}.

To investigate the role of stochasticity on the final native state,
an inverse strategy is proposed, based on a simple angular 3D
random-walk model to build up protein backbones with different
lengths and distinct percentages of secondary structures. In the
proposed model, each step has a fixed radial size $l_0$ but dihedral
$\Phi$ and $\Psi$ angles of the protein backbone are chosen
according to independent Gaussian probability distributions,
following the suggestion given in reference \cite{shaw03}. The mean
value and standard deviation of each defined according to the
allowed regions of the $\Phi/\Psi$ plot of the frequency
distribution of dihedral angles, the so called Ramachandran map.
$\Phi$ and $\Psi$ mean values were used as proposed in the PRELUDE
software package \cite{rooman91}. These values were computed from
comparative statistics of the backbone secondary structure for
several amino acid sequences. Table \ref{table} indicates the seven
possible pairs of $(\Phi,\Psi)$ dihedral angles and the associated
structures of the main chain backbone, as predicted by this method.
These specific angles describe the average conformation of a wide
range of proteins with known backbone structures.
\begin{table}[ht]
\par
\begin{center}
\begin{tabular}{|c|c|c|}
\hline $\Phi$ & $\Psi$ & Conformation \\
\hline -65 & -40 & $A$ \\
\hline -89 & -1 & $C$ \\
\hline -117 & 142 & $B$ \\
\hline -69 & 140 & $P$ \\
\hline  78 & 20 & $G$ \\
\hline  103 & -176 & $E$ \\
\hline  -83 & 133 & $O$ \\
\hline
\end{tabular}
\end{center}
\caption{Seven possible pairs of dihedral angles and the associated
conformations occurring in several amino acid sequences
\cite{rooman91}. The $\alpha$-helix pair is denoted by $A$ while the
$\beta$-strands pair is denoted by $B$.}  \label{table}
\end{table}
To simulate structures with  a definite percentage of secondary
structures $f$, a characteristic number of steps $n$ is fixed and
the growth process within these $n$ steps is divided in
two stages:\\
1) The first $n \times f$ steps are built accordingly to an angular
Gaussian probability distribution, whose mean value is one pair of
angles as seen in Table\,\ref{table}, which in turn is associated
with a given structure. \\
2) The next $n \times (1-f)$ steps are built according to an angular
Gaussian probability distribution, whose mean value at each step is
randomly chosen from amongst the seven pairs of angles of Table
\ref{table}.\\

For the following $n$ steps rules $1$ and $2$ are repeated
upwards to construct a peptide chain with $N$ amino acids.
Therefore, in order to obtain an appropriate choice of the $f$
percentage this stochastic procedure  assures  that the final
peptide main chain follows the Ramachandran map.

Within this simple model  structures of the protein backbone were
constructed  considering only the dihedral angles.  All other
bonded or non-bonded interactions were not explicitly considered as well as
excluded volume and steric effects, which are expected to
be taken into account by the appropriated choice of the average values of the
Gaussian probability distributions. For this reason it was
possible  to generate  an elevated  number of samples of possible
protein conformations.  An exhaustive number of simulations ($\sim
10^4$) were performed  considering three basic cases: (a) $f=0.6$
with $\alpha$-helix  structures; (b) $f=0.6$ with $\beta$-strand
structures and (c) the first $ n/2 $ steps built with $f=0.6$ of
$\alpha$-helix  structures and the next $n/2$ steps with $f=0.6$
of $\beta$-strand  structures, consecutively. Therefore, in the
$\alpha$-case (a) $60\%$ of the amino acids corresponds  on
average to $\alpha$-helix structures, in the $\beta$-case (b)
$60\%$ of the amino acids corresponds to $\beta$-strands, while in
the mixed-case (c) the whole structure has an average of  30$\%$
of $\alpha$-helix and 30$\%$ of $\beta$-strands. $10^4$ chains of
the  total size varying from $N=125$ to $400$, with the number of steps
 $n=100$ were generated. For each case described above there was
a variation of  $f$ in the interval $[0,1]$, step $0.1$. There was
also a variation of the standard deviation $\sigma$ of the
Gaussian distribution within a wide range of values from $0$ to
$\pi$.

\begin{figure}[ht]
\begin{center}
\includegraphics[width=10cm]{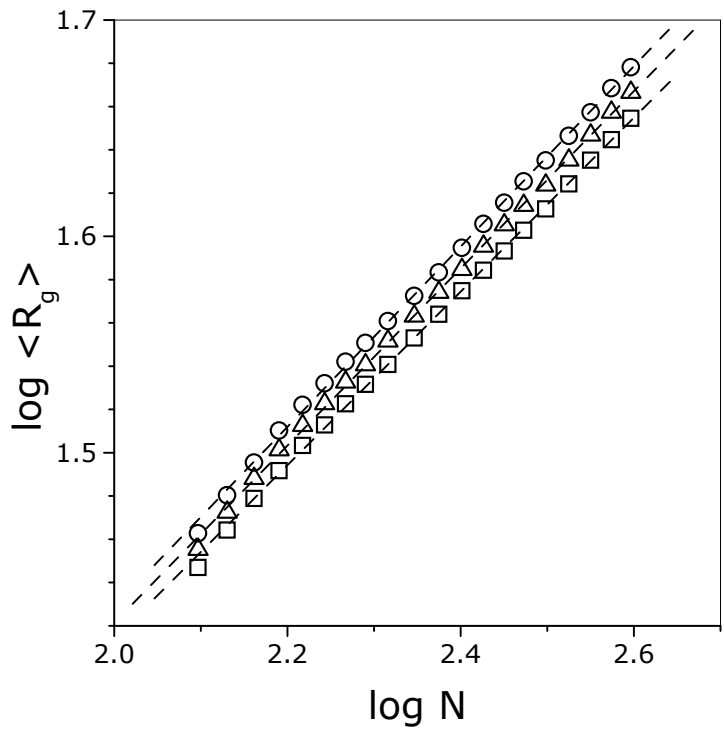}
\end{center}
\caption{The \emph{average} radius of gyration  in function of the number
of residues obtained from the simulations with $f=0.60$ for the
$\alpha$-helix structures ($\square$), the mixed structures
($\triangle$) and $\beta$-strands structures ($\bigcirc$) with
scaling exponent $0.401 \pm 0.002$, $0.409 \pm 0.002$ and $0.417 \pm
0.002$, respectively, obtained from the relation $<R_g>\sim N^\nu$.
Each point results from the average of $10^4$ simulations.
The dashed lines  indicate  the linear fitting. The  error  bars are
smaller than the symbol sizes.  In all cases $\sigma/\pi=0.1$ were
fixed. }\label{figure1}
\end{figure}

Figure  \ref{figure1}  shows  the average radius of gyration ($<
R_{g}>$)  in function of  the number of amino acids ($N$) for the
three distinct  choices  of structures. From this plot
 however, a power-law behavior pattern can be observed
indicating that these structures are self similar. The corresponding
scaling exponent, which somehow describes the compactness of the
structure,  is  calculated by the scaling relation: $R_{g} \sim
N^{\nu}$, in all cases. The characteristic scaling exponents are
$\nu = 0.401 \pm 0.002$ for the $\alpha$-helix case and $\nu = 0.417
\pm 0.002$ for the $\beta$-strands case,  which falls in the
interval to values of the real proteins. For the mixed case
$\nu=0.409\pm 0.002$  were achieved.
To further analyze the compactness of structures based on the
$\alpha$-helix, the  $\beta$-strand and that  composed of a mix of
$\alpha$-helices and $\beta$-strands the scaling exponent $\nu$ was
estimated in function of the percentage of secondary structure $f$.

\begin{figure}[ht]
\begin{center}
\includegraphics[width=10cm]{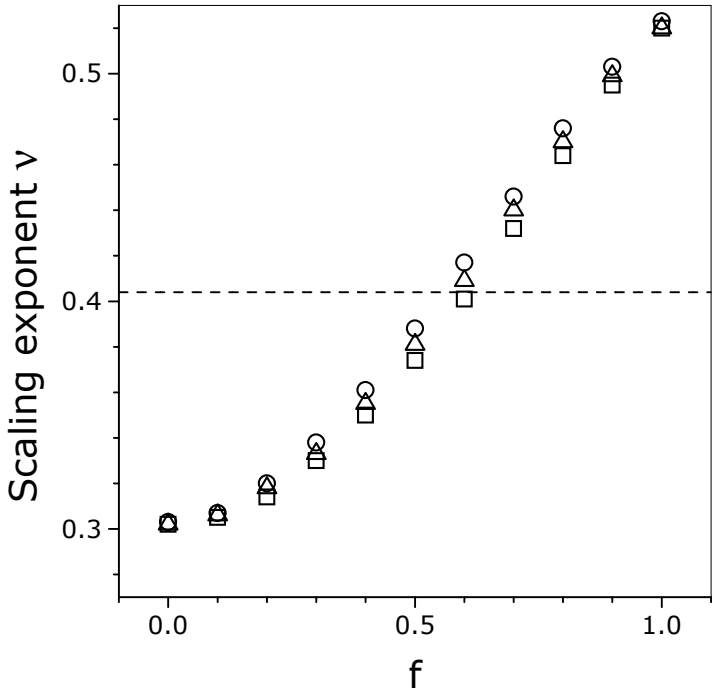}
\end{center}
\caption{The dependence of the scaling exponent $\nu$  in  function
of the percentage $f$ of secondary structures, for $\alpha$-helices
($ \square$), $\beta$-strands ($\bigcirc$) and the mixed cases
($\triangle$). The  error  bars are smaller than the symbol sizes.
In all cases $\sigma/\pi=0.1$  was fixed.  The dashed line indicates
the experimental value $\nu \simeq 0.405$.}\label{figure2}
\end{figure}

\begin{figure}[ht]
\begin{center}
\includegraphics[width=10cm]{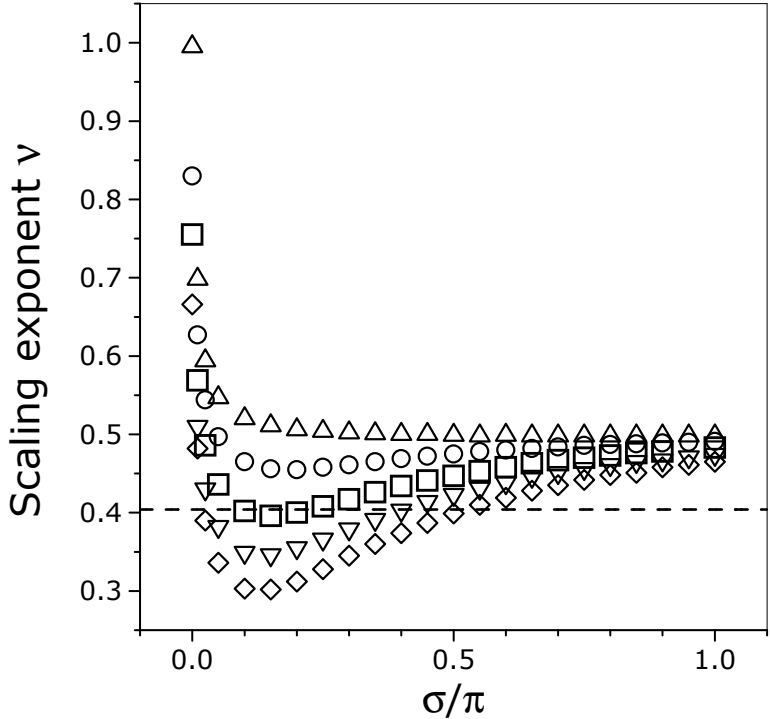}
\end{center}
\caption{The dependence of the scaling exponent $\nu$  in   function
of the variance $\sigma$ of the Gaussian probability distributions
of the dihedral angles (in units of $\pi$), for the $\alpha$-helix
structure, considering the value of $f=0\,(\lozenge)$,
$f=0.40\,(\triangledown)$, $f=0.60\,(\square)$, $f=0.80\,(\bigcirc)$
and $f=1.0\,(\vartriangle)$. The dashed line indicates the
experimental value $\nu \simeq 0.405$.}\label{figure3}
\end{figure}

In the Figure \ref{figure2} it can also observed that for
the three studied cases the scaling exponent $\nu$ growth is from
$\nu=0.302\pm0.002$ when the percentage of secondary structures is
close to zero ($f \simeq 0$), i.e., a complete random structure,
up to $\nu_{\max}=0.520\pm0.001$ corresponding to full ordered
structures. Slight different values was observed within the
interval $0 < f < 1$ for the structures composed of
$\alpha$-helices, $\beta$-strands and the mixed case, the one
built with $\alpha$-helix motifs being more compacted than the
other cases. It is worth to mention that the lower limit for the
scaling exponent is $1/3$, which would correspond to a fully
compact three-dimensional structures commonly observed for
globular proteins. However, the plots of Figure \ref{figure3}
shows that the scaling exponent lies below $1/3$ up to $\sim 0.3$,
for lower values of the fraction of secondary structure motif ($0
\leq f<0.3$). Hence, such interval will corresponds to a high
fraction $(1-f)$ of a random structure -- built with dihedral
angles chosen from Gaussian distributions centered at values
randomly chosen at each step from the values given in Table I.
Therefore, it is expected that the excluded volume and steric
effects will not play their role and the model fails to reproduce
the expected $\nu=1/3$ limit behavior.

At this point  the behavior of the exponent $\nu$ against changes
in the Gaussian probability distributions variance is explored.
The increased dispersion of the $(\Psi,\Phi)$ Gaussian
distribution corresponds to the increase of randomness in the
chain structure, destroying the role of the $f$ percentage of
secondary structures.   Figure  \ref{figure3} illustrates the
behavior of $\nu$ in function of $\sigma/\pi$ of the
$\alpha$-helix case considering the value of
$f=0.0,\,0.40,\,0.60,\,0.80$ and $1.0$.   For elevated variance
values an increase of the $\nu$ exponent approaching to $0.5$ was
observed for all values of $f$, which is associated to the lack of
ordering. On the opposite limit, for vanishing variance values an
increase of the $\nu$ exponent was observed toward to one, which
can be easily proved to correspond to the power law exponent of
the linear chain backbone structure for the $f=1.0$ case. However,
between these two limits a minimum value of $\nu$ is observed
around $\sigma/\pi=0.15$, independent of the value of $f\,(\neq
1.0)$, corresponding to the maximum compacted structures.
Therefore, when fitted with the $\nu$ value extracted from data,
as discussed below, the corresponding optimum value of $f$ is
found to be around $f\simeq 0.60$.

\begin{figure}[ht]
\begin{center}
\includegraphics[width=10cm]{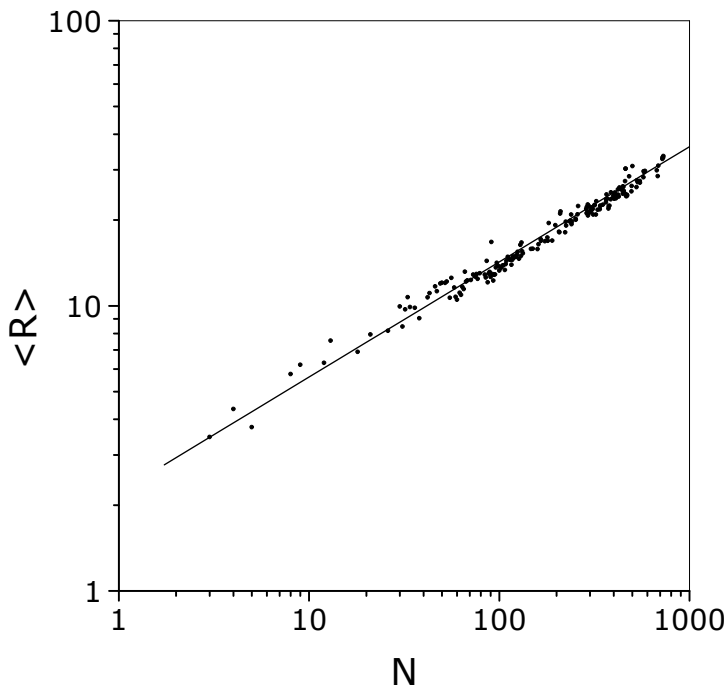}
\end{center}
\caption{The behavior of the average radius $<R>$ against the number
of amino acids $N$ for a set of 1826 proteins chains. The scaling
exponent $\nu = 0.40 \pm 0.02$. }\label{figure4}
\end{figure}

1826 different protein chains deposited in the Brookhaven Protein
Data Bank (PDB) were also investigated in order to provide a
comparison with the obtained simulation results. The number of amino
acids was measured in function of the average radius. Figure
\ref{figure4} depicts the main characteristics of all systems
discussed herein using geometric (dihedral angles) analysis. The
figure indicates that several protein chains deposited in the PDB
have a self-similar behavior pattern when the average radius ($<R>$)
is plotted against the number of amino acids ($N$). In this case the
average radius signifies the average distance from the geometric
centre of all coordinates \cite{moret05a}. It was also noticed that
an average value was calculated for radii of chains with the same
number of amino acids. This intrinsic characteristic of the protein
structures must be responsible for explaining several aspects of
these molecules such as the high compactness, which has been
discussed in several other different contexts
\cite{dill90,wolynes95,richards74,liang01,bagci02,moret05b,moret06,moret05c}.
The $\nu$ exponent associated with the average radius obtained is
$0.404 \pm 0.008$, which is in agreement with recent similar study
involving 200 proteins \cite{enright05}. The volume and mass of
proteins with more than fifty amino acids scale with respect to the
average radius with exponents $\delta=2.47\pm 0.04$ \cite{liang01}
and $\delta=2.47\pm0.03$ \cite{moret05b} respectively. This
consequently corresponds to an exponent ($\nu=1/\delta \approx
0.405$) if we assume that the mass scales to the average radius with
exponent one. This exponent would be associated to mixed chains
composed of secondary structures, which according to the present
simulations vary in an interval of $0.401 \leq \nu \leq 0.417$.
Furthermore,  the PDB proteins presents  structures with $ \sim$ 60
$\%$ of secondary structures \cite{voet95}, which  justifies  and
confirms  our initial assumption of $f=0.6$ for simulated structures
shown in Figure \ref{figure1}. It is worth mentioning  that the
present model can be generalized for  growth chains with distinct
percentages of different structures, accordingly  to  the
corresponding protein Ramachandran map.

Through  the approach presented herein  the protein folding
problem has been investigated assuming that proteins fold in a
mixed manner following some directed process while being subject
to certain stochastic ingredients. This signifies that the process
is neither completely random, as raised by the Levinthal's
paradox, nor it is entirely driven by the physical chemistry
principles that establish a definite folding pathway. The simple
model  presented herein focuses on the stochastic aspects of the
formation of the secondary structures which is believed to be the
earliest relevant precursor event in the folding, as  confirmed by
other  recent experimental evidence \cite{honig99}. According to
rules 1 and 2 of the present model  it can be assumed  that  the
formation of a secondary structure ($\alpha$-helix or
$\beta$-strands) occurs during a consecutive fraction $f$ of steps
governed by a Gaussian probability distribution, whose parameters
(mean and standard deviation) are extracted from data associated
with the Ramachandran map. These parameters, which  caracterize  a
given structure, reflect  the physical and chemical  processes
underlying  protein stability. Therefore this fraction $f$ of the
chain somehow mimics the interplay between energy stability and
entropy. Thus  to the extend that  the structure reaches a certain
size it  looses stability and folds  randomly, changing the mean
value of the Gaussian probability distribution at  each  step, but
still using the possible values extracted from data.

What emerges from this stochastic process is that the
narrow Gaussian probability distribution of helical or stranded
arrays do provides  an insight into  the protein folding process,
which goes beyond the possibilities of molecular structure or
molecular dynamics analysis. Actually, this narrow Gaussian
probability distribution supplies a peptide backbone chain with
self-similar properties that matches with the one estimated from
experimental data (see Figure \ref{figure4}). Furthermore, Figure
\ref{figure3} illustrated that if a probability distribution with
wide values of the variance (large values of $\sigma/\pi$) is
considered approaching to an uniform probability distribution, the
resulting final structure was less compacted than that obtained
with the Gaussian distribution process. One further interesting
result obtained by the present model is that backbone chains with
$\alpha $-helix motifs are more compact than the $\beta$-strands
and the mixed case, a result confirmed by current literature
\cite{dill90,dill93,dill97}. The fractal dimension ($\delta =
1/\nu \approx $ 2.49) obtained from Figure \ref{figure1} and
Figure \ref{figure4} are comparable with  that obtained by the
volume analysis against radius \cite{liang01} or by mass-size
exponent analysis ($\delta \approx 2.47$)
\cite{moret05b,moret06,moret05c,enright05}.

The  results  of this study indicate that simulated structures are
more compact when secondary portions ($\alpha $-helices  and/or
$\beta $-sheets)  are present, than those  built with other  sets of
dihedral angles, as shown  in Table \ref{table}.  The method was
systematically compared  to  other widely used methods of protein
folding analysis \cite{dill90,wolynes95,moret98,wales99,zhou99a}.
Several of these methods do not result in a fully consistent
assignment of self-similarity of protein structures. It should be
mentioned the recent work of Huang \cite{huang07} where a
sophisticated conditioned self-avoid walk model was proposed taking
into account the hydrophobic effect and the hydrogen bonding
focusing on the physical chemistry mechanisms underlying the protein
folding process. Independent of the details of the underlying
physical chemistry mechanisms, building protein backbones with the
method proposed in the present work suggests that these structures
are driven by narrow Gaussian distributions. Thus it is the general
conclusion of this work that protein folds like an angular
Gaussian-walk and not as a random-walk problem.

\end{document}